\documentclass[11pt]{article}

\usepackage{geometry}  
\usepackage{subcaption} 
\usepackage[T1]{fontenc}  
\usepackage[utf8x]{inputenc}  
\usepackage{fancyhdr}  
\usepackage{enumitem}  
\usepackage{hyperref}  
\usepackage{titling}  
\usepackage{natbib}  
\usepackage{mathtools}  
\usepackage{titlesec}  
\usepackage{lastpage}  

\usepackage[english]{babel}  
\usepackage{amsmath}  
\usepackage{amsfonts}  
\usepackage{amssymb}  
\usepackage{wasysym}  
\usepackage{bbm}  
\usepackage{array}  
\usepackage{xr}  
\usepackage{verbatim} 
\usepackage{float} 
\usepackage{authblk}

\usepackage{suffix}

\newcommand{\INLINEBOX}[2]{%
   \begin{center}%
    \fcolorbox{#1!60!black}{#1}{%
      \addtolength{\linewidth}{-0.6cm}
      \begin{minipage}{\linewidth} #2 \end{minipage}%
    }%
   \end{center}\vspace{1pt}%
}

\newcommand{\MARGINBOX}[1]{%
  \mbox{}%
  \marginpar%
   [\tiny\raggedleft\hspace{10pt}#1]%
   {\tiny\raggedright\hspace{0pt}#1}%
}

\usepackage{color}

\newcommand{\TODO}[2][]{\MARGINBOX{\textcolor{red}{\emph{ToDo (#1):}} #2}}
\WithSuffix\newcommand\TODO*[2][]{\INLINEBOX{red!20!white}{\emph{ToDo (#1):} #2}}

\newcommand{\FIXME}[2][]{\MARGINBOX{\textcolor{blue!80!black}{\emph{FixMe (#1):}} #2}}
\WithSuffix\newcommand\FIXME*[2][]{\INLINEBOX{blue!20!white}{\emph{FixMe (#1):} #2}}

\newcommand{\NOTE}[2][]{\MARGINBOX{\textcolor{green!80!black}{\emph{Note (#1):}} #2}}
\WithSuffix\newcommand\NOTE*[2][]{\INLINEBOX{green!20!white}{\emph{Note (#1):} #2}}

\hypersetup{%
  pdfborderstyle={/S/U/W 1}
}

\setcitestyle{square,numbers,sort,unsrtnat}
\titleformat{name=\section}{\normalfont\Large\bfseries}{}{0pt}{}
\titleformat{name=\subsection}{\normalfont\large\bfseries}{}{0pt}{}
\titleformat{name=\subsubsection}{\normalfont\normalsize\bfseries}{}{0pt}{}

\newcommand{\keywords}[1]{\textbf{Keywords}: {#1}}

\newcommand{\optincludegraphics}[2][]{}

\newcommand{\optinput}[1]{}

\newtagform{brackets}{[}{]}
\usetagform{brackets}

\newcommand{\thejournal}[1]{Phys.\Rev.\ AB}

\newcommand{\V}[1]{\mathbf{#1}} 
\newcommand{\opal}{\textsc{OPAL}}

\newcommand {\RM}[1]{\mathrm{#1}}
\renewcommand{\Re}{\mathbb{R}}

\title{Intensity limits of the PSI Injector II cyclotron}

\begin{document}

\begin{titlepage}

\author[1,2,3] {A. Kolano}
\author[2] {A. Adelmann}
\author[3] {R. Barlow}
\author[2] {C. Baumgarten}

\affil[1] {CERN, CH-1211 Geneva 23, Switzerland}  
\affil[2] {Paul Scherrer Institut, 5232 Villigen PSI, Switzerland}
\affil[3] {The University of Huddersfield, Huddersfild HD1 3DH, UK}
\date{25th July 2017}
\maketitle

\begin{abstract}

We investigate  limits on the current of the PSI Injector  II high intensity separate-sector isochronous cyclotron, in its present configuration and after a proposed upgrade. Accelerator Driven Subcritical Reactors, neutron and neutrino experiments, and medical isotope production all  benefit from increases in current, even at the $\sim 10\%$ level: the PSI cyclotrons provide relevant experience.  As space charge dominates at low beam energy, the injector is critical. Understanding space charge effects and halo formation through detailed numerical modelling gives clues on how to maximise the extracted current.
Simulation of a space-charge dominated low energy high intensity  (9.5 mA DC) machine, with a complex collimator set up in the central region shaping the bunch, is not trivial.
We use the OPAL code, a tool for charged-particle optics calculations in large accelerator structures and beam lines, including 3D space charge.
We have a precise model of the present (\textit{production}) Injector II,  operating at 2.2 mA current.  A simple  model of the proposed future (\textit{upgraded}) configuration of the cyclotron  is also investigated.
 We estimate intensity limits  based on the developed models, supported by fitted scaling laws and measurements. 
We have been able to perform more detailed analysis of the bunch parameters and halo development than any previous study. Optimisation techniques enable better matching of the simulation set-up with Injector II parameters and measurements.
We  show that in the production configuration the beam current scales to the power of three with the beam size. However, at higher intensities, 4th power scaling is a better fit, setting the limit of approximately 3 mA. Currents of over 5 mA, higher than have been achieved to date, can be produced if the collimation scheme is adjusted. 
\end{abstract}


\end{titlepage}
\pagebreak

\section{Introduction}

There are few parameters that can be adjusted to increase the beam intensity of an existing cyclotron at constant losses. The most promising is to increase the turn separation by increasing the accelerating voltage. A different strategy to follow, more difficult to realize, is to develop a compact and stationary distribution.

Since the beam radius of a centered cyclotron beam is a monotonic function of energy, a higher accelerating voltage reduces the number of turns and leads to larger radial increments between turns. For extraction by an electrostatic deflector a large turn separation is required to  reduce beam losses at the septum, which is the main source of beam losses in Injector II. Turn separation at extraction can, within limits, also be increased by appropriately shaping the radial profile of the magnetic field in the extraction region and/or by the use of a controlled betatron oscillation. This has already been implemented in injector II~\cite{privM}.

Injector II is equipped with two main RF cavities operating at about 400 kV, and two 30 kV third-harmonic cavities which were originally used to produce a  flat-top but are now operated in accelerating phase. The nominal extraction energy is 72 MeV. Injector II delivers up to 2.4 mA to the PSI Ring cyclotron, which accelerates the beam to 590 MeV and provides up to 1.4 MW beam power to the targets in the experimental hall.
Injector II has been in routine operation since 1985 and the performance has been improved significantly over the years. A major step forward in terms of beam intensity was made in the mid 1990s by the discovery of the space charge dominated beam operation without flat-topping~\cite{Adam:0,Adam:1}.  Since then, the third-harmonic cavities are used in accelerating mode, further decreasing the number of turns.

The next planned upgrade of Injector II involves replacing the third-harmonic cavities with new high voltage resonators. This will decrease by 20 the number of turns needed to reach 72 MeV, and nearly double the turn separation at extraction.
In order to achieve optimal performance with the new resonators it is useful to study the parameters that determine bunch and halo formation beforehand to check whether the collimator locations can be improved and to determine the optimal settings. For this study, a 3D beam dynamics model has been developed for Injector II in its present production configuration and for the machine after the upgrade, and the results of this model are presented here.
%
\section{Injector II}

\begin{figure}[!h]
   \centering
   \includegraphics*[width=110mm]{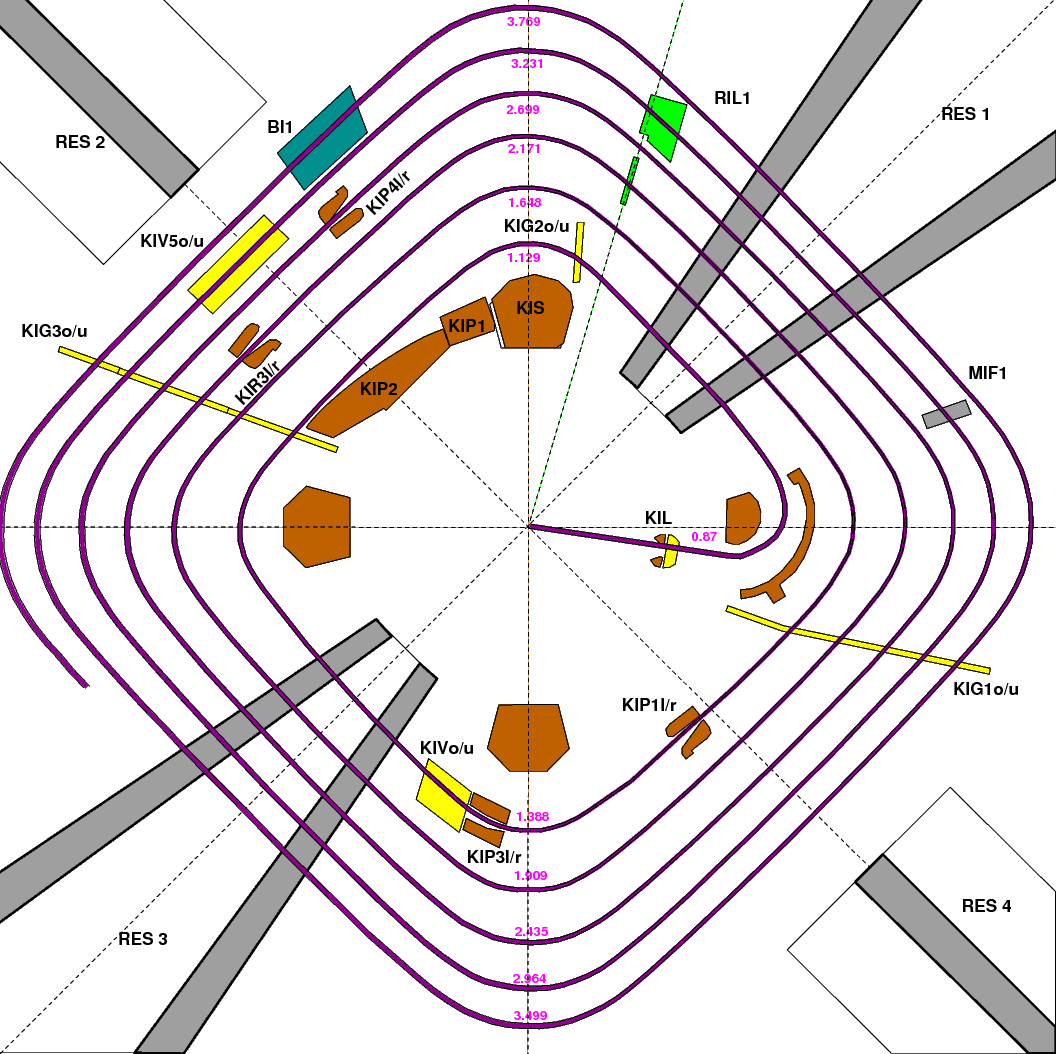}
   \caption{Central region of Injector II showing the collimators.\ Vertical ones are highlighted in yellow and horizontal in orange. }
   \label{inj2coll}
\end{figure}

A Cockcroft-Walton preinjector provides a 12 mA DC proton beam of 870 keV, which is prebunched and then vertically injected into the center of Injector II~\cite{Olivo1992}. As shown in Figure~\ref{inj2coll},  several vertical and radial collimators control the beam current and shape the beam within the first 5 turns at energies below 3.7 MeV, leading to excellent beam properties~\cite{Stammbach2001}. In particular the KIP1 and KIP2 collimators cut away 2.4 and 4.8 mA respectively for standard operation  at  2.2 mA. 

Injector II is a 72 MeV high-intensity machine dominated by space charge effects, particularly in the central region.  Generally  such effects cause beam deterioration, however Injector II exhibits a special feature: strong transverse and longitudinal coupling, combined with space charge, creates a ``vortex" inside the bunch with spiralling tails that wrap around the compact core.  Round compact bunches are produced after these tails are removed by collimators in the first turns~\cite{Stammbach1996}.

Injector II operates in its present {\em production} configuration with two double-gap resonators located $180^\circ$ apart, and two single-gap 3rd harmonic cavities used as additional accelerating cavities.\ The machine accelerates from 870 keV to 72 MeV in 83 $\pm$1 turns.\ In this study we use the values for the voltages on the cavities used on a typical day in 2015, and shown in Table~\ref{table:1}.\
 
\begin{table}[h!]
\caption{RF cavities parameters. }
\centering
\begin{tabular}{c c  }
\hline\hline
Name & Peak Voltage (kVp) \\ [0.5ex] 
\hline
Resonator 1 (gaps RF0 and RF1) &431.357\\
Resonator 2 (gaps RF2 and RF3) &400.374\\ 
Flat-top 3 (RF4)& 31.698\\
Flat-top 4 (RF5)& 31.382\\[1ex]
 
\hline
\end{tabular}
\label{table:1}
\end{table}
 Replacement of the two 3rd harmonic cavities by two single-gap 400 kV cavities is part of planned upgrade that will enable production of higher intensity beams~\cite{bopp_hipa}.\ Extrapolating using a $V^{3}$ scaling law~\cite{joho81} predicts that to deliver 3 mA currents after the upgrade, an energy gain of about 1.2 MeV per turn is needed. Acceleration will then take place in only 60 turns, and the turn separation is expected to nearly double at extraction.

\section{Methods and Models}

\subsection{The electrostatic particle-in-cell method used in \opal}

We consider the
Vlasov-Poisson description of the phase space, including external and
self fields.  Let $f(\mathbf{x},\mathbf{v},t)$ be the density of the particles in the
phase space, i.e.\ the position-velocity $(\mathbf{x}, \mathbf{v}) \in \Re^{3}\times\Re^{3}$
space and $t > 0 \in \Re$ denotes the time.  Its evolution is determined by the collisionless Vlasov
  equation,
\begin{equation*} \label{eq:Vlasov}
  \frac{df}{dt}=\partial_t f + \mathbf{v} \cdot \nabla_{\mathbf{x}} f
  +\frac{q}{m}(\mathbf{e}+ \mathbf{v}\times\mathbf{b})\cdot
  \nabla_{\mathbf{v}} f  =  0,
\end{equation*}
where $m$, $q$ denote particle mass and charge, respectively.  The
electric and magnetic fields $\mathbf{e}$ and $\mathbf{b}$ are
superpositions of external fields and self fields (space charge),
\begin{equation}\label{eq:allfield}
    \mathbf{e} =
    \mathbf{e^{\RM{ext}}} + \mathbf{e^{\RM{self}}}, \qquad
    \mathbf{b} =
    \mathbf{b^{\RM{ext}}} + \mathbf{b^{\RM{self}}}.
\end{equation}


If $\mathbf{e}$ and $\mathbf{b}$ are known, then each particle can be
propagated according to the equation of motion for charged particles in an
electromagnetic field. After particles have moved, we have to 
update $\mathbf{e^{\RM{self}}}$ and $\mathbf{b^{\RM{self}}}$ (among other things).  
For this, we change the coordinate system into the one moving with the
particles.  By means of the appropriate Lorentz
  transformation $\mathcal{L}$~\cite{lali:84} we arrive at a (quasi-) static
approximation of the system in which the transformed magnetic field
becomes negligible, $\hat{\mathbf{b}}\! \approx\! \mathbf{0}$.  The
transformed electric field is then obtained from
\begin{equation}\label{eq:e-field}
  \hat{\mathbf{e}}=\hat{\mathbf{e}}^{\RM{self}}=-\nabla\hat{\phi},
\end{equation}
where the electrostatic potential $\hat{\phi}$ is the solution of the
Poisson problem
\begin{equation}\label{eq:poisson0}
  - \Delta \hat{\phi}(\mathbf{x}) =
  \frac{ \mathcal{L}(\rho(\mathbf{x}))}{\varepsilon_0},
\end{equation}
equipped with appropriate boundary conditions.  Here, $\rho$ denotes the spatial charge
density and $\varepsilon_0$ is the dielectric constant.
By means of the inverse Lorentz transformation ($\mathcal{L}^{-1}$) the electric field
$\hat{\mathbf{e}}$ can then be transformed back to yield both the
electric and the magnetic fields in~\eqref{eq:allfield}.

The Poisson problem~\eqref{eq:poisson0} discretized by finite
differences can  be solved efficiently on a rectangular grid by a
Particle-In-Cell (PIC) approach~\cite{qiry:01}.  The right hand side
of~\eqref{eq:poisson0} is discretized by sampling the particles at the
grid points.  In~\eqref{eq:e-field}, $\hat{\phi}$ is interpolated at the
particle positions from its values at the grid points. 

\subsection{Equations of motion}

We integrate in time $N$ identical particles, all having the rest mass $m$
and charge $q$.  The relativistic equations of motion for particle $i$
are
\begin{align}
  \frac{\mathrm{d}\V{x}_i}{\mathrm{d}t} &= \frac{\V{p}_i}{m \gamma_i}, \\
\frac{\mathrm{d}\V{p}_i}{\mathrm{d}t} &= q \left(\V{e}_i+\frac{\V{p}_i}{m \gamma_i} \times \V{b}_i\right), \label{eq:dpdt}
\end{align}
where $\V{x}_i$ is the position, $\V{p}_i = m \V{v}_i \gamma_i$ the relativistic momentum, $\V{v}_i$ the velocity, $c$ the speed of light and the Lorentz factor 
$\gamma_i = 1 / \sqrt{1-(||\V{v}_i||/c)^2} = \sqrt{1+(||\V{p}_i||/(mc))^2}$. The electric and
magnetic field, $\V{e}_i$ and $\V{b}_i$, can be decomposed into external field and self field contributions:
\begin{align}
\V{e}_i &= \V{e}^{\mathrm{ext}}(\V{x}_i, t) + \V{e}^{\mathrm{self}}(i, \V{x}_{1 \ldots N}, \V{p}_{1 \ldots N}), \\
\V{b}_i &= \V{b}^{\mathrm{ext}}(\V{x}_i, t) + \V{b}^{\mathrm{self}}(i, \V{x}_{1 \ldots N}, \V{p}_{1 \ldots N}).
\end{align}
The notation $\V{x}_{1 \ldots N}$ is a shorthand for $\V{x}_1, \ldots,
\V{x}_N$, and is used for other vectors analogously. The self field describes the
field created by the collection of particles i.e. the source of the Coulomb repulsion.
The external electromagnetic field (from magnets etc.), which can have an explicit
dependence on time $t$, are in this model treated as independent of the other particles.

\subsection{Three models of Injector II}

\begin{figure}[h] 
   \centering
   \includegraphics[width=\textwidth]{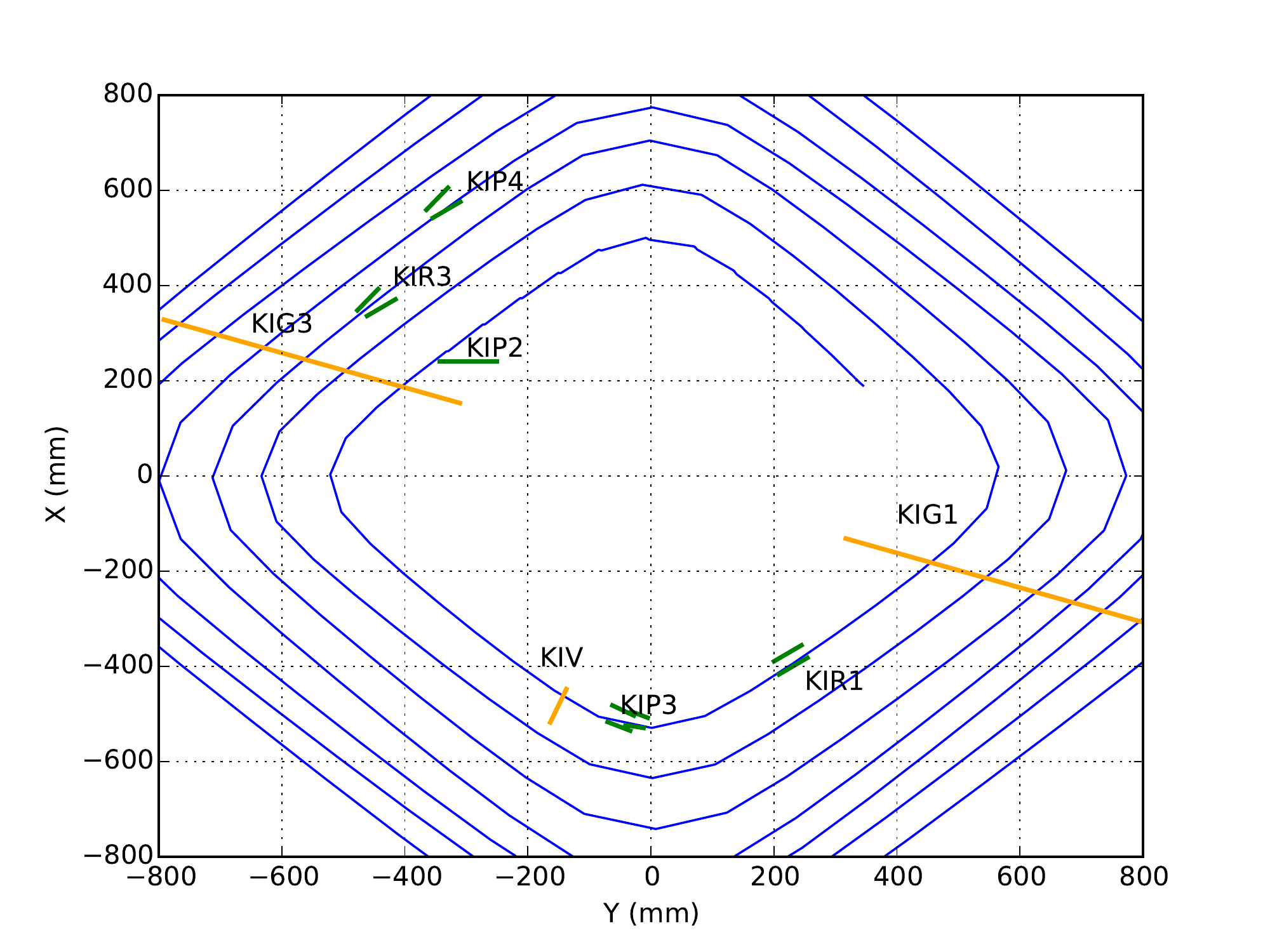}
   \caption{Modelled central region of the Injector II with marked (green) horizontal (yellow) vertical collimators.}
   \label{collimators}
\end{figure}

In order to investigate the limits on the beam current, three models of the
collimation system in Injector II have been developed, for both the production and the upgraded configuration.\  The models, which have been validated and analysed for various intensities, statistics and Injector II measurements, are: 
\begin{description}
\item [4$\sigma$ cut.]  During simulation of the acceleration, the effect of the collimators is modelled by continually imposing a 4$\sigma$ cut in the $x$, $y$ and $z$ planes.  This model can give an estimate of RMS (Root Mean Square) beam sizes and emittance, but not the halo.\  Despite its simplicity, the results show an impressively matched beam, accelerated up to 72 MeV\cite{AnnaThesis}.

The full beam extent is around 20 mm, close to that measured in Injector II.
\item [6-turn 4$\sigma$ cut.] 
Removing particles outside the 4$\sigma$ region throughout the acceleration will not give a full picture of the halo formation.\ As one of the ongoing questions is whether halo is re-formed after the collimators, some simulations have been performed with the beam being cut with 4$\sigma$ only for the first 6 turns, i.e. in the region where the collimators are positioned\cite{AnnaThesis}. 
\item [Physical collimation.]
For detailed halo studies the exact effects of the collimators on the beam are  simulated. 
The simulation starts at a $30^\circ$ angle from the centre of  the SM1 magnet: the central $90^\circ$ injection is not included in the model as the central region beam is strongly correlated and generating bunches that would represent the real beam in the machine is not trivial.
Also several collimators such a KIP1 and KIS2, that protect the magnets or cavities, were not included 
as they have no effect on normal beams.  At the point where we start the simulation, an 11 mA bunch is cut by KIS2 and KIP1 collimators, therefore 9.5 mA is believed to be a good approximation to the beam current at this point. 
Figure~\ref{collimators} shows the modelled central region of Injector II with the collimators included in the simulations.\ 
\end{description}

The precise beam dynamics model is developed using the OPAL (Object Oriented Parallel Accelerator Library) C++ based simulation code~\cite{opal}. It is a tool for charged-particle optics calculations in large accelerator structures and beam lines including 3D space charge.\ 

Injected, initially uncoupled, bunches of particles undergo linear deformations/transformations under the influence of fields generated by the beam optics and space charge as they travel through an accelerator.\ Particle-in-cell (PIC) simulations require generation of such multivariate particle distributions.\ We used multivariate Gaussian particle distributions with covariances which are theoretically matched to a closed orbit of the injector cyclotron~\cite{Baumgarten2011,Baumgarten2012}. However, the presumptions of the theoretical model are hardly fulfilled in the first turns of the Injector cyclotron: The beam is not coasting (not even approximately), the energy gain is large compared to the kinetic energy and the RF field is not weak.

 Table~\ref{table:2} gives the main parameters used in the production mode model
 
\begin{table}[h!]
\caption{Simulation parameters. }
\centering
\begin{tabular}{c c c }
\hline\hline
Name & Parameter value& Unit\\ [0.5ex] 
\hline
Turn number & 83$\pm$1 & \\
Injection energy &868.5 & keV\\ 
Energy& 72 &MeV\\
Radius& 392.0 &mm\\
Resonator gap voltage&215.7 &kV$_{peak}$\\
3rd harmonic& 31.0& kV$_{peak}$\\
Added voltage offset& 87.919& kV$_{peak}$\\
RF phase& $50^\circ$& degrees\\
Azimuth&$30^\circ$& degrees\\
Radial momentum&  -5.50008 $\cdot10^{-3}$ &$\beta \gamma$\\[1ex]
 
\hline
\end{tabular}
\label{table:2}
\end{table}

\section{Validation with measurements}

\begin{figure}[h]  
   \centering
   \includegraphics[width=\textwidth]{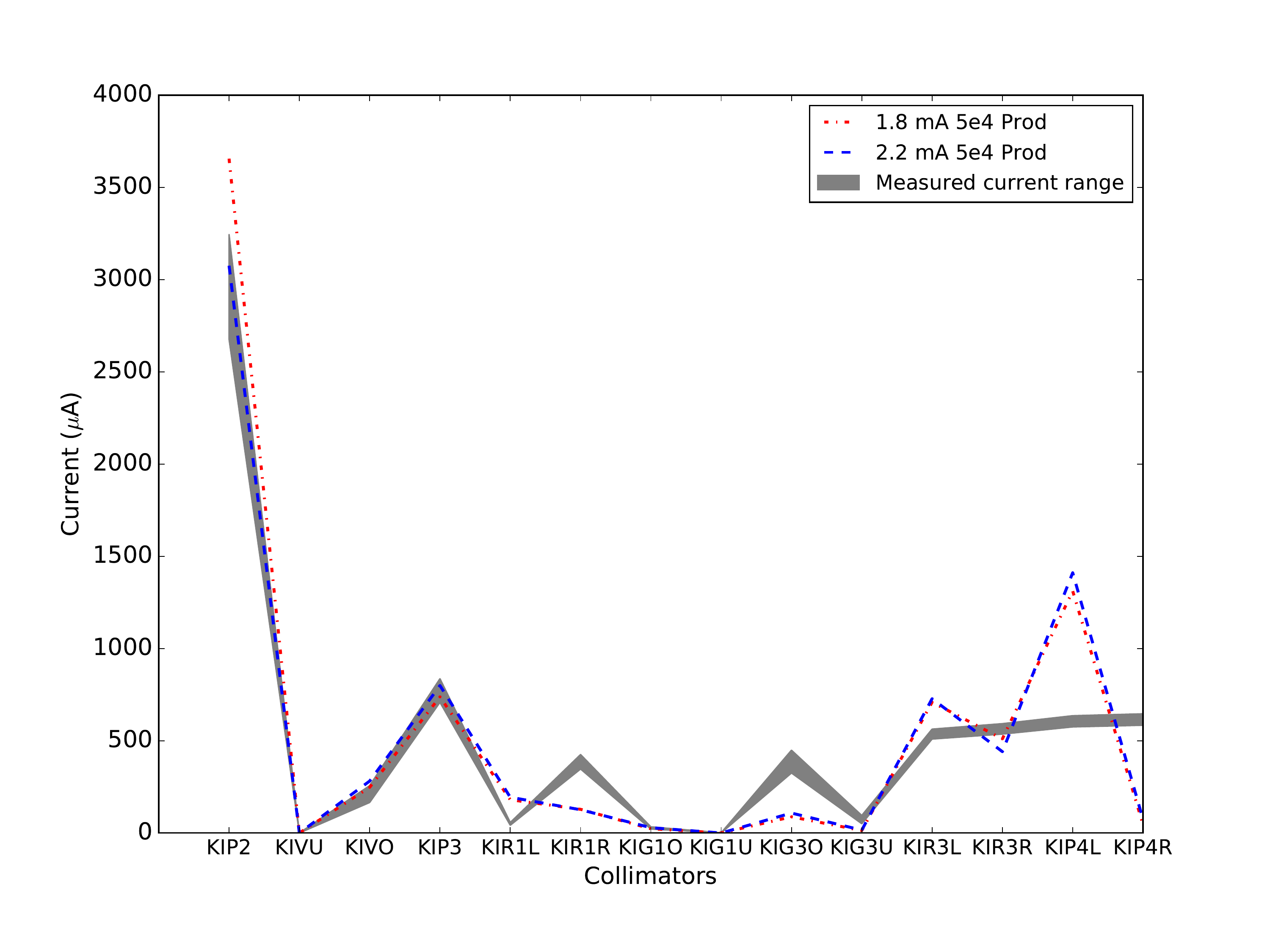}
   \caption{Current measured on the collimator (L-left, R-Right) leaves  in Injector II compared to simulations.  The measured current  is shown in grey, simulations for a 1.8 mA and 2.2 mA beam current in red and blue respectively.  }
   \label{MesCol}
\end{figure}

We have expanded the 3D model to include actual collimators and probes, and validated this set-up with measurements.\ Figure~\ref{MesCol} shows the extracted currents measured on the collimators, taken over one day (which gives a 5\% spread), compared to the simulated ones for a $\sim$2 mA beam current. The main limitation  on the simulations is the small number of macroparticles
used, but as Injector II has been in use over several decades, 
many parameters are not known with great precision, and some values will change with each commissioning. Although most collimator positions in Injector II are considered to be well known, some,  in particular KIP4,  could have an error of as much as a  few centimeters. Also, although the magnetic field map is considered to be accurate, it may not include a full description of the  trim coil fields. 
The currents measured on collimators are intended only as an indication of losses and not for accurate particle flow  measurement.
Therefore some
mismatch in simulated versus measured data is expected, and the comparison is  only to give a general idea of the collimator position and  aid simulation set-up.

Nevertheless agreement is good for most collimators, though there appears to be a discrepancy in the results from KIP4 Left and Right this is not surprising (for reasons explained earlier) and the sum is consistent.  The 5\% discrepancy in measurement can be due to equipment limitations and secondary electron emission~\cite{priv}.

\begin{figure}[h] 
   \centering
   \includegraphics[width=\textwidth]{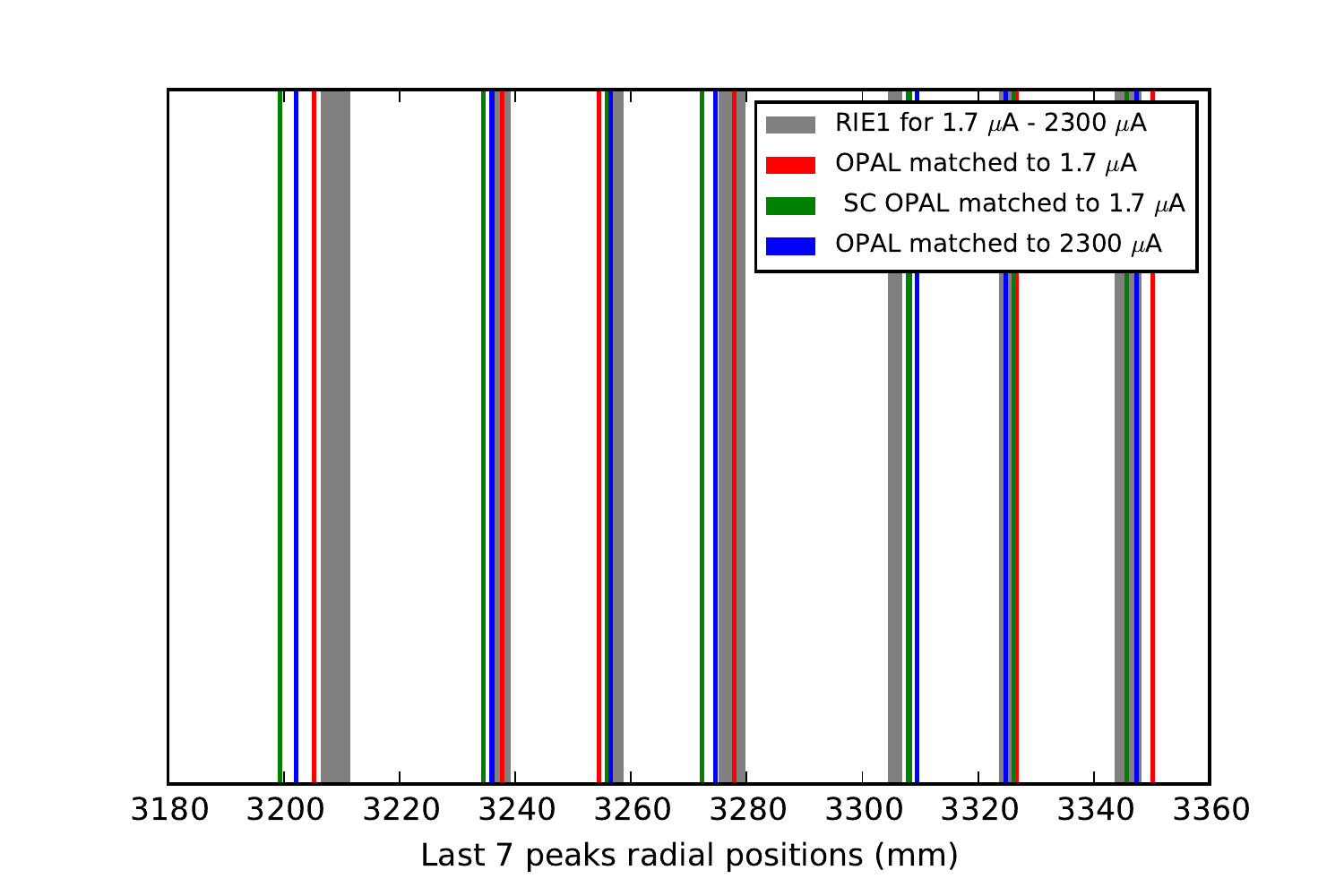}
   \caption{Positions of measured and simulated radial intensity peaks of Injector II in the production configuration.\ Measured peak locations are in grey, the simulated peaks matched to the lower current bound of 1.7 mA are in red and those  the higher 2.3 mA bound in blue.\ The peak positions of multiparticle space-charge run are in green.}
   \label{peakMatch}
\end{figure}

For further validation, we use measurements of the radial position of the turns at the highest energies, as measured by  the RIE1 radial probe, which is located at $158^\circ$ from the SM1 magnet.\ This probe has been introduced in the model, detecting radial intensity peaks.  A single particle optimisation has been used to search for initial conditions that will match the last 7 radial intensity peaks. We have successfully found starting conditions that produce orbit pattern matching   on a sub millimeter level.  Results are shown in Figure~\ref{peakMatch}. The Injector II turn pattern changes with intensity, thus there is no fixed radial peak positions but rather a range, due to both the trim coil adjustment and tune manipulation in the last turns.  This is a very sensitive quantity to simulate correctly, as  it requires an accurate description over all turns in the acceleration. Simulations were done for single particles assuming currents of 1.7 and 2.3 mA, and also including space charge at 1.7 mA.  
The good match of measured and simulated radial intensity peaks achieved is remarkable.

\section{Beam size and Profile Parameter}

\begin{figure}[!h]
 \hspace*{-1.0cm}
   \centering
   \includegraphics*[width=160mm] {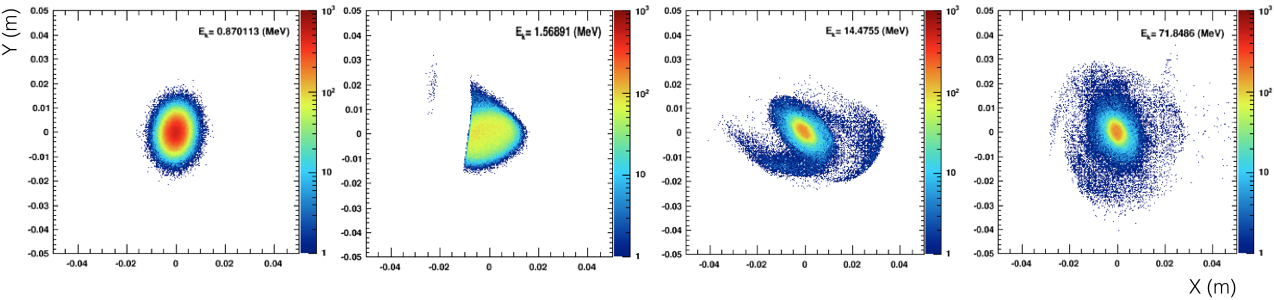}
  \caption{An example of progression of the distribution in the configuration space passed through collimators accelerated from 0.87 to 1.57, 14.48, and finally  72 MeV.\ Despite significant distribution deformations due to large cuts (the second snapshot is directly after KIP2)  the tails ``wrap" around the spiralling centre of the distribution leading to formation of a stable core with halo around.}
   \label{relpln}
\end{figure}

The  analysis of the simulated distribution passed through the collimators and accelerated to 72 MeV indicates that even though the bunch is significantly cut by the collimators (KIP2 alone cuts around 50\%), we observe the formation of a steady compact core with some halo around it (Figure~\ref{relpln}). This is due to the strong longitudinal-transverse coupling, combined with space charge.

To quantify the halo, we have chosen to use a simple kurtosis-like parameter $\it{h}$, applied to 1D spatial projections of the distribution~\cite{wangler}.\  It is defined as
\begin{equation}  
h = \frac{\langle x^4\rangle}{\langle x^2 \rangle^2} - 2 .
\end{equation}

It is just the 4th central moment, scaled by the square of the second moment. The subtraction of 2 means that $h=1$, for a Gaussian distribution. While this definition is based on a continuous beam,
Wangler and Crandall provided a normalisation constant for bunched beams. The constant $15/7$,
for example, 
would make $h=0$ for a uniform density bunch in $x,y,z$ space, for details we refer to  \cite{Allen2001}.

The excess above one is a sign of the presence of a significant tail~\cite{Allen2001, Allen2002}, hence a development of halo. We have chosen this particular definition of the halo parameter because it can be related most easily to measurements. These measurements can be taken at
positions in the machine where $\langle x p_{x}\rangle = 0$ and thus where our profile parameter is equivalent to the halo parameter
of other authors~\cite{Allen2001}.

\begin{figure}[!h]
\centering
   \begin{subfigure}[b]{0.45\textwidth}
   \includegraphics[width=1\linewidth]{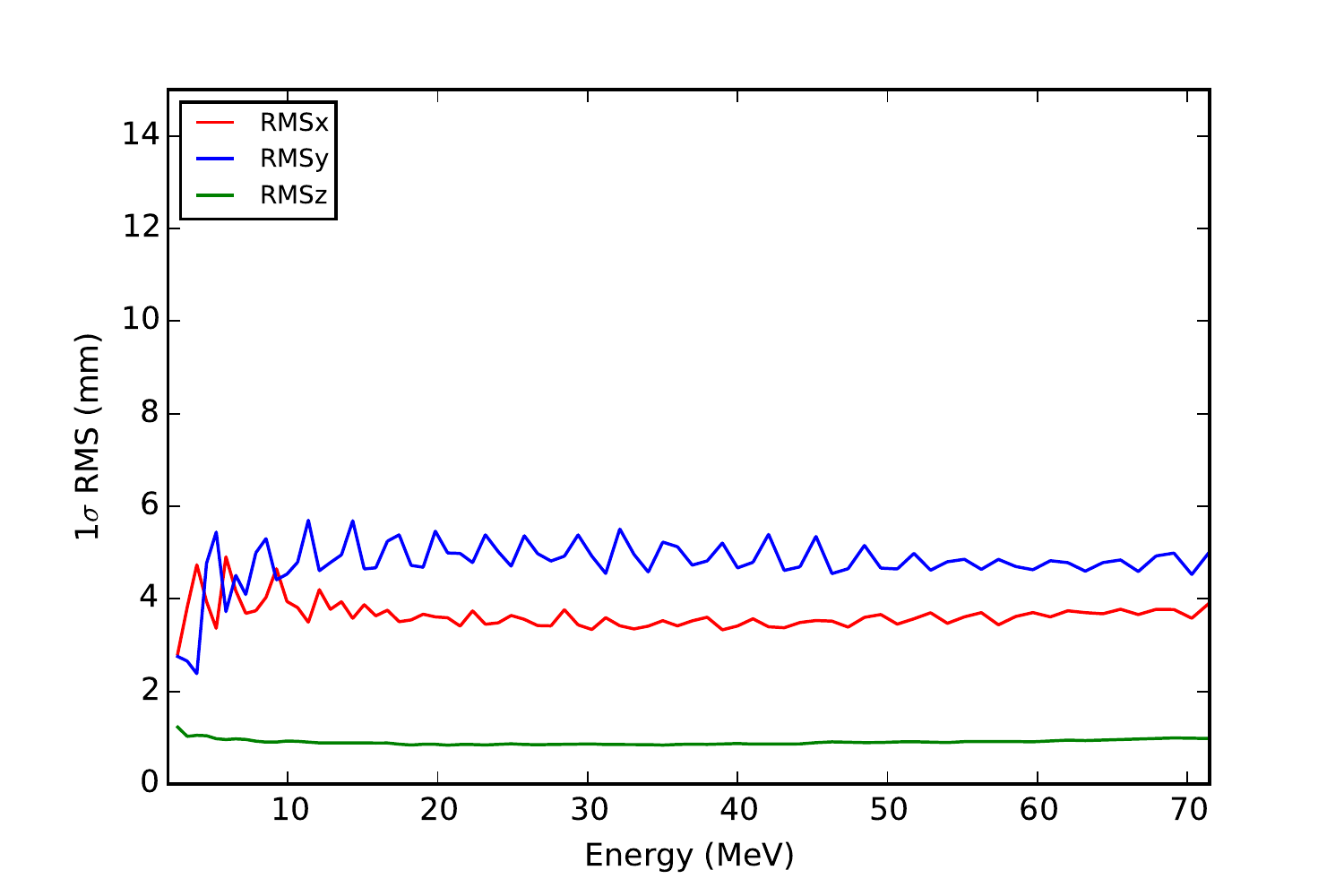}
   \caption{RMS beamsize}
   \label{fig:rms} 
\end{subfigure}
 ~ 
\begin{subfigure}[b]{0.45\textwidth}
   \includegraphics[width=1\linewidth]{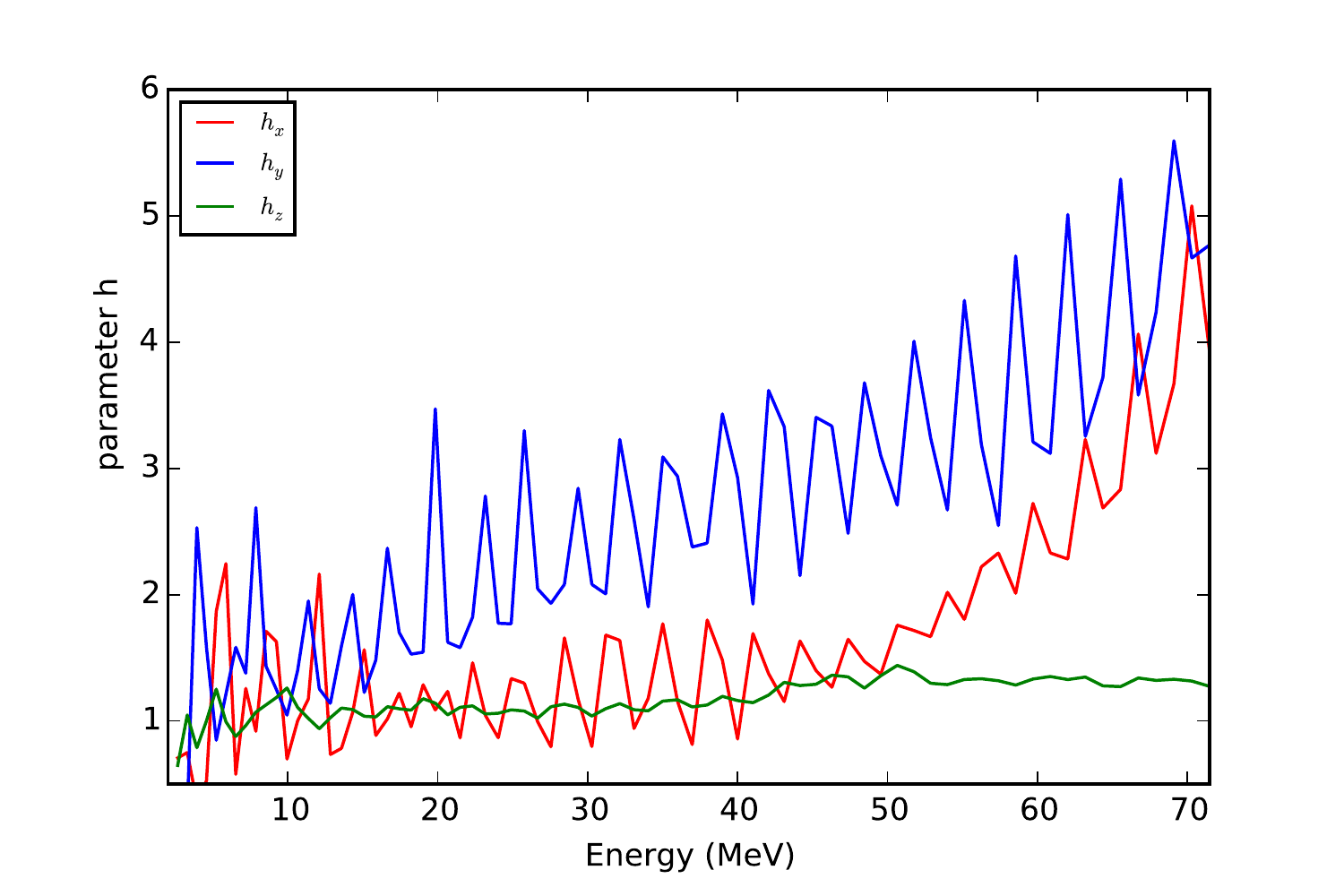}
   \caption{h parameter}
   \label{fig:h}
\end{subfigure}

\caption{Development of the rms beam size (a) and the halo parameter (b) for the production configuration at 2.3 mA beam current, using the  6-turn 4$\sigma$ cut. The rms size is stable in $x$, $y$ and $z$,  However after about 40 turns a horizontal halo forms, due to the the longitudinal plane coupling to the radial plane.}
 \label{fig:rmsh}
\end{figure}

In Figure~\ref{fig:h} we observe halo formation around the stable core shown in Figure~\ref{fig:rms}, as the beam accelerates.

\section{Limits of beam current in  Injector II}

 We have analysed and compared the predictions of all three collimation models for both the production and the upgraded configurations.\  The beam-current limit depends on the turn separation at the the extraction septum; for the real machine this is 20 mm, centre to centre. Simulations suggest this could be increased by appropriate parameter manipulation.\ 
 
\begin{figure}[h]
 \centering
   \includegraphics*[width=140mm] {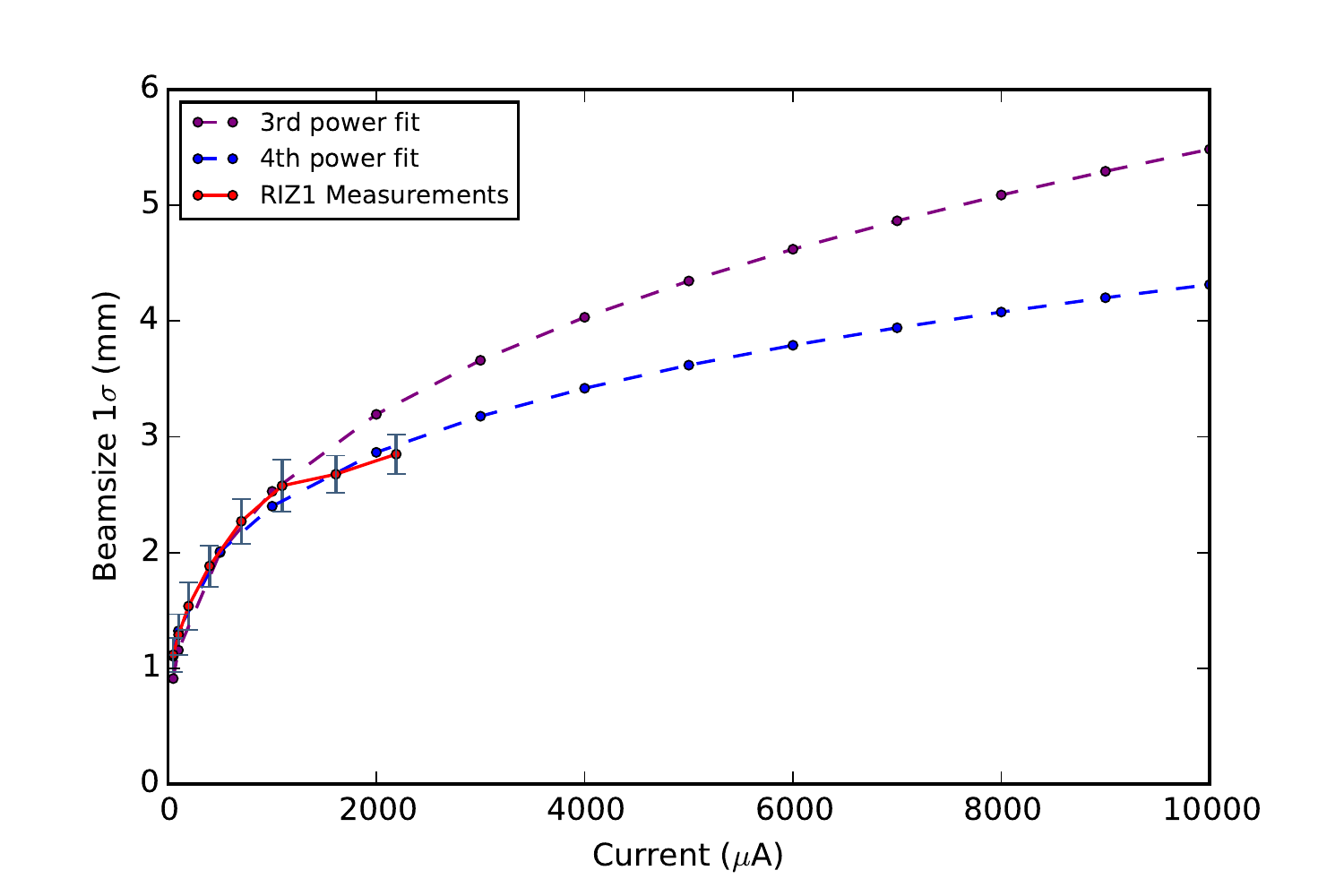}
  \caption{Measured beam size (red) compared to 3rd (purple) and 4th (blue) power scaling.}
   \label{scaleYN}
\end{figure}

The RMS beam size can be found experimentally by measurements using the RIZ1 probe, which is  located on the extraction orbit.  This is shown for various beam currents in Figure~\ref{scaleYN}.

We denote the final  RMS beam size by  $\sigma$.
  It can be considered as an intrinisic beam size
  $\sigma_{0}$ and a contribution $\sigma_{sc}$,
  the blow up (or hopefully shrinking) due to space charge,
  strong longitudinal-transverse coupling and lattice 
  nonlinearities. 
\begin{equation}
\sigma^{2} = (\sigma_{0} + \sigma_{sc})^{2},
\end{equation}

we note that the expected beam size $\sigma$ of a spherically symmetric space charge dominated cyclotron beam is expected to obey~\cite{bertrand,Baumgarten2011}
\begin{equation}
\sigma^4=k_1\,I\,\sigma+k_2\,\varepsilon^2\,.
\end{equation}
with two constants $k_1$ and $k_2$ that are independent of current and emittance.
If the first term on the right dominates, we speak of space charge domination and expect  the beam size to scale as the cube root of the beam current:
\begin{equation}
\sigma \propto \root 3 \of {I} .
\end{equation}
However, if the second term cannot be neglected, then the exact scaling law depends on the relationship between emittance and current. If the emittance is  large and constant one would expect a 4th power scaling at low currents. But in our case, since the beam current in Injector II is controlled by the opening of KIP 2, there are non-trivial correlations between beam current and emittance, both horizontal and vertical. We therefore cannot make a precise theoretical prediction.

The general form of the beam size scaling can be written as:
\begin{equation}
\sigma \propto a \sqrt[n]{I}-b,
\end{equation}
where $a$, $b$ and $n$ $\in$ ${\rm I\!R}$.A

We therefore try empirical fits for both $n=3$ and $n=4$, the 3rd and 4th order. These are also shown in Fig.~\ref{scaleYN}. 
Although the $\chi^{2}$  of 1.32   for the 3rd root scaling is acceptable, a  beam size depending on the current like the 4th root gives a  a better fit, particularly for higher currents:  $\chi^{2}$ in this case is 0.83. (Both values are actually very small, for 6 degrees of freedom, suggesting that the quoted errors on the measurements have been overestimated). 

For the cube-root scaling, the   1$\sigma$ RMS beam size depends on the current as:

\begin{equation}
\sigma \propto 0.2\sqrt[3]{I}+ 0.4, 
\end{equation}

whereas the fourth-root  scaling gives:
\begin{equation}
\sigma \propto 0.44 \sqrt[4]{I}-0.06. 
\end{equation}

\begin{figure}[h]
 \centering
      \includegraphics*[width=160mm] {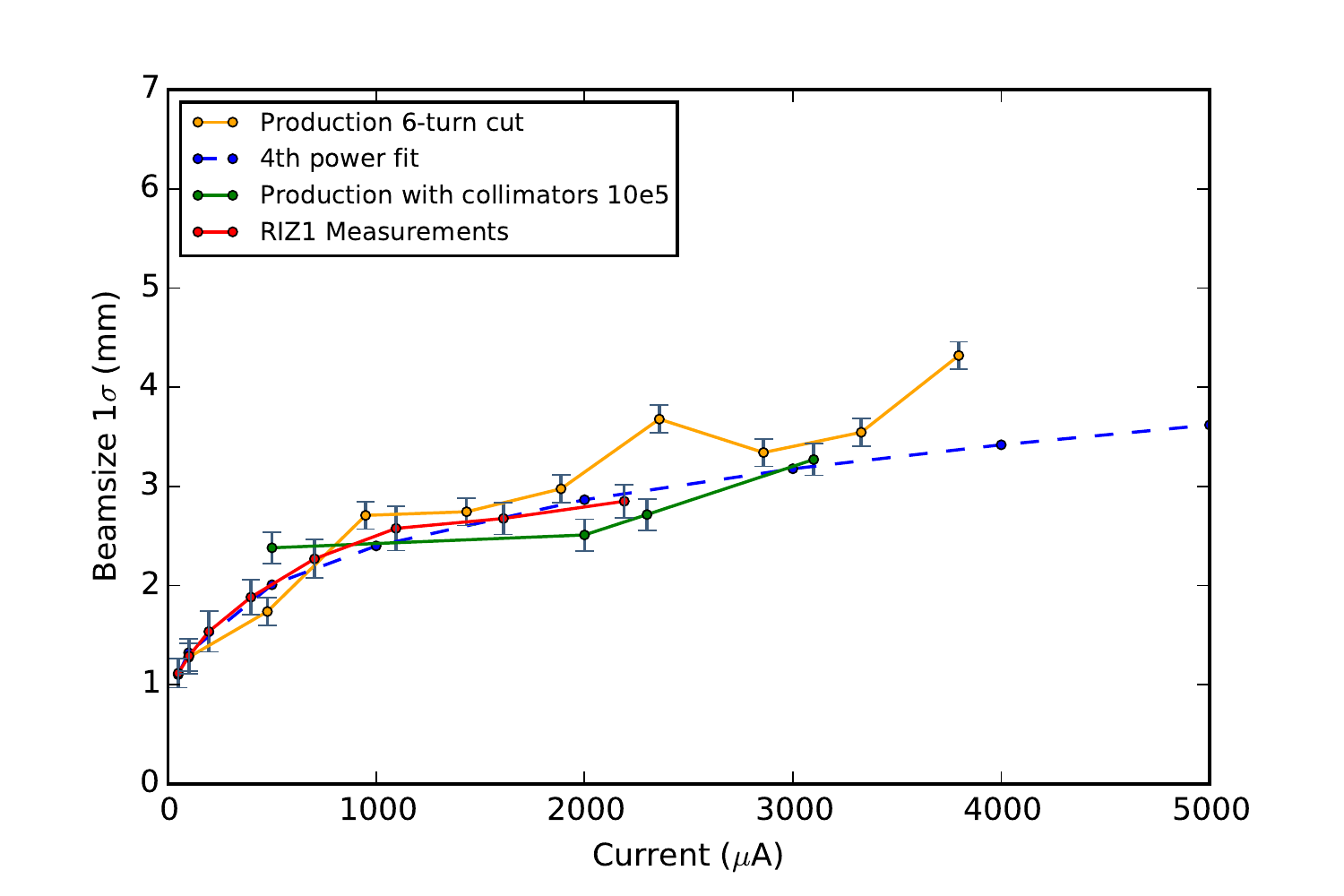}
  \caption{6-turn 4$\sigma$ and (yellow) and physical collimator (green) model simulation results compared to measurements(red)  and new suggested scaling (blue).}
   \label{scalelogYN}
\end{figure}

In view of the limited experimental data, we also consider the predictions of our simulations. These are shown in 
Figure~\ref{scalelogYN}. The beam size of the 6-turn model matches the measurements well up to 2 mA, as does that of the physical collimator model. If, indeed, the bunch size scales like the 4th root, the difference from the 3rd root prediction appears only at currents above 2 mA. At higher beam currents the predictions of the $1-\sigma$ beam size broadly agree with the 4th root fit, the prediction of the 6-turn 4-$\sigma$ model may may go slightly over, but not by much.  The beam size stays below $\sim$ 3 mm up to high beam currents which, given the 20mm beam separation,   implies  that higher intensities, of order of 3 mA, could be achieved with the existing production configuration.

\begin{figure}[h] 
   \centering
   \includegraphics[width=\textwidth]{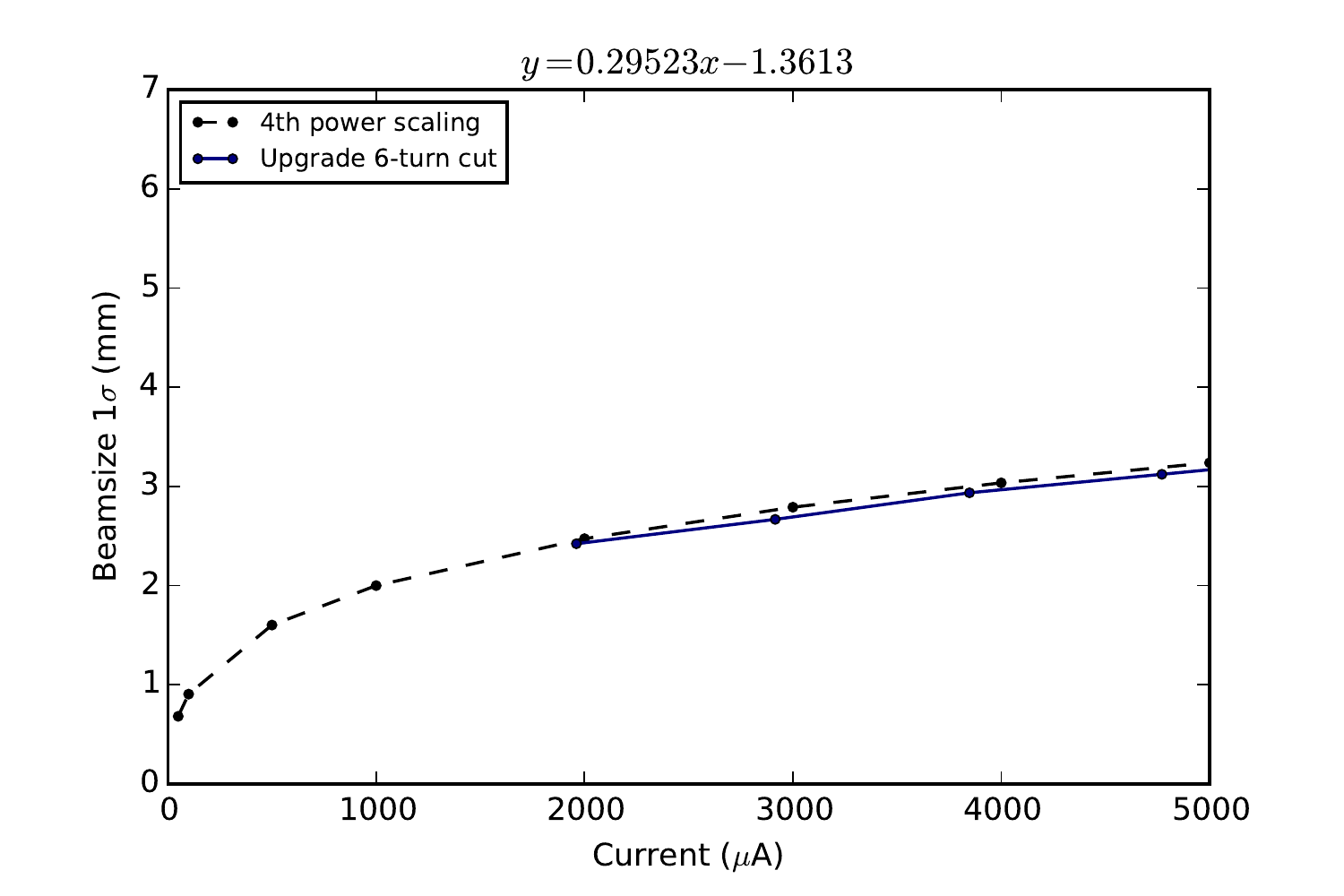}
   \caption{Plot of beam size evolution with intensity of the upgraded 6-turn 4? model with its fits}
   \label{newScale}
\end{figure}

 We then explore the upgraded configuration (for which only simulation data is, of course, available) using the 6-turn 4$\sigma$ cut model. Fitting simulated data to the fourth root,, we see in Figure~\ref{newScale} that the $1-\sigma$ RMS beam size is related to the current as follow:
\begin{equation}
\sigma \propto 0.44 \sqrt[4]{I}-0.5.
\end{equation}

The cube-root scaling becomes:

\begin{equation}
\sigma \propto 0.17 \sqrt[3]{I}+0.3 .
\end{equation}

$\chi^{2}$  is 0.11 and 0.079 respectively, which is not a significant enough difference, but both parameterisations and the simulation indicate that a beam current of  5 mA could be reached, with an RMS beam size of only 3.2 mm.

  For benchmarking, and for the estimation of intensity limits, horizontal data is sufficient as it is the width of the septum that is the main physical constraint, although large longitudinal tails will also decrease the limits.\  Based on the measurements and simulations we have attempted to find scaling laws that govern the beam size variation with intensity, and  thus intensity limits.\ We have found that the production configuration current scales to the power of four with the beam size, setting the limit to approximately 3 mA.\  Further analysis of an upgraded configuration suggests that intensities of over 5 mA could be produced with an adjusted collimation scheme.\

\section{Conclusions}

The Injector II separate-sector isochronous cyclotron has been in operation for over 30 years, initially designed to provide up to 1 mA beam current~\cite{Schryber1975}.
It is the only cyclotron worldwide known to take advantage of the cyclotron specific correlation for space charge dominated beam transport. Beams of 2.2 mA are routinely extracted.

In this works, three models of Injector II have been developed for both the production and upgraded configuration:  continuous 4$\sigma$ cut, 6-turn 4$\sigma$ cut,  and real physical collimation.\  These models have been validated and analysed with data from radial profile measurements (RIZ1), and from loss rates read from the collimators.

 Measured and simulated data, up to $1600~ \mu A$, are in good agreement with the simplified theoretical model which predicts a beam size that scales with the cube root of the beam current.\ However, measurements and simulations of higher currents fit a fourth root variation better.\ If this conjecture holds, Injector II could operate at 3 mA with its current resonators and flat-top cavities.\  This limit nearly doubles for the planned upgrade, where the currents could reach over 5 mA.\

This study suggests that that the Injector II has not yet reached its ultimate limit. With the aid of a precise beam dynamics model, an improved scaling law could be found.  A matched distribution generator, even though it only uses linear space charge, was used to obtain approximative initial conditions.\ The upgraded set-up looks promising with nearly doubled turn separation and intensity.\  

\bibliography{refsA1} 
\end{document}